%% file: main.tex
\documentclass[conference]{IEEEtran}
\usepackage{ifpdf}

\usepackage{cite}

\ifCLASSINFOpdf
   \usepackage[pdftex]{graphicx}
   \graphicspath{{../pdf/}{../jpeg/}}
   \DeclareGraphicsExtensions{.pdf,.jpeg,.png}
\else
   \usepackage[dvips]{graphicx}
   \graphicspath{{../eps/}}
   \DeclareGraphicsExtensions{.eps}
\fi

\usepackage{amsmath,nccmath}
\usepackage{amsthm}

\usepackage{algorithmic}

\usepackage{array}
\usepackage{multirow}

\usepackage{stfloats}

\usepackage[moderate,title]{savetrees}

\ifCLASSOPTIONcaptionsoff
  \usepackage[nomarkers]{endfloat}
 \let\MYoriglatexcaption\caption
 \renewcommand{\caption}[2][\relax]{\MYoriglatexcaption[#2]{#2}}
\fi

\usepackage{url}

\usepackage[absolute,overlay]{textpos}
\usepackage{units}
\usepackage{amssymb}
\usepackage{cleveref}
\usepackage{dsfont}
\usepackage{soul}
\usepackage{comment}
\usepackage{caption}
\usepackage{matlab-prettifier}
\usepackage{pdfpages}
\usepackage{subcaption}
\usepackage{tikz,pgfplots}
\usetikzlibrary{positioning}

\newtheorem{thm}{Theorem}
\newtheorem{prob}[thm]{Problem}
\newcommand\blfootnote[1]{%
  \begingroup
  \renewcommand\thefootnote{}\footnote{#1}%
  \addtocounter{footnote}{-1}%
  \endgroup
}

\begin{document}

\title{\Huge{\textnormal{System-level thermal and electrical modeling of battery systems for electric aircraft design}}}

\author{Thomas Kuijpers, Jorn van Kampen and Theo Hofman}

\maketitle
\IEEEpeerreviewmaketitle
\blfootnote{The authors are with the Control Systems Technology (CST) section, Dep. of Mechanical Engineering, Eindhoven University of Technology (TU/e), Eindhoven, 5600 MB, The Netherlands, \texttt{t.hofman@tue.nl}}


\input{Sections/Abstract}

\input{Sections/Introduction}
\input{Sections/Methodology}

\input{Sections/Results}

\input{Sections/Conclusion}

\section*{Acknowledgment}
\noindent We thank Dr.~I.~New for proofreading this paper. This paper was partly supported by the NEON research project (project number 17628 of the Crossover program which is (partly) financed by the Dutch Research Council (NWO)).

\bibliographystyle{IEEEtran}
\bibliography{ReferencesThomas_offline}

\clearpage
\thispagestyle{plain}

\ifCLASSOPTIONcaptionsoff
  \newpage
\fi

\end{document}

%% file: Sections/Abstract.tex
\begin{abstract}
This work introduces a framework for simulating the electrical power consumption of an 8-seater electric aircraft equipped with high-energy-density NMC Lithium-ion cells. We propose an equivalent circuit model (ECM) to capture the thermal and electrical battery behavior.
Furthermore, we assess the need for a battery thermal management system (BTMS) by determining heat generation at the cell level and optimize BTMS design to minimize energy consumption over a predefined flight regime. The proposed baseline battery design includes a 304-kWh battery system with BTMS, ensuring failure redundancy through two parallel switched battery banks. Simulation results explore the theoretical flight range without BTMS and reveal advantages in increasing battery capacity under specific conditions. Optimization efforts focus on BTMS design, highlighting the superior performance of water cooling over air cooling. However, the addition of a 9.9 kW water-cooled BTMS results in a  16.5\% weight increase (387 kg) compared to no BTMS, reducing the simulated range of the aircraft from 480 km to 410 km. Lastly, we address a heating-induced thermal runaway scenario, demonstrating the robustness of the proposed battery design in preventing thermal runaway.
\end{abstract}

%% file: Sections/Introduction.tex
\section{Introduction}\label{sec:introduction}

Electric aviation, driven by emission reduction goals like EU ACARE Flightpath 2050 and NASA's N+3 program, has promising potential to lower the aviation industry's carbon footprint \cite{Adu-Gyamfi2022ElectricTechnologies}. However, current technological limitations restrict large electric passenger aircraft to light and short-distance flights \cite{Epstein2019ConsiderationsPropulsion}. To address this, companies are exploring small-scale, short-distance passenger aircraft using advanced High Energy Density (HED) Lithium-ion battery (LIB) technology. LIBs of the `pouch' type are preferred for their efficient packing and capacity \cite{Plett2015BatteryModeling}. Increasing battery capacity extends flight range but adds weight, affecting energy consumption and design optimization. Safety concerns, especially maintaining optimal battery temperature, are critical \cite{Yang2021EssentialVehicles}. A battery thermal management system (BTMS) is essential for safety and efficiency. This study proposes a minimum-energy framework to manage heat generation in a pouch-type battery system for an 8-seater electric aircraft, ensuring it stays within thermal limits.


\begin{figure}[t]
    \centering
    \includegraphics[width=0.85\linewidth]{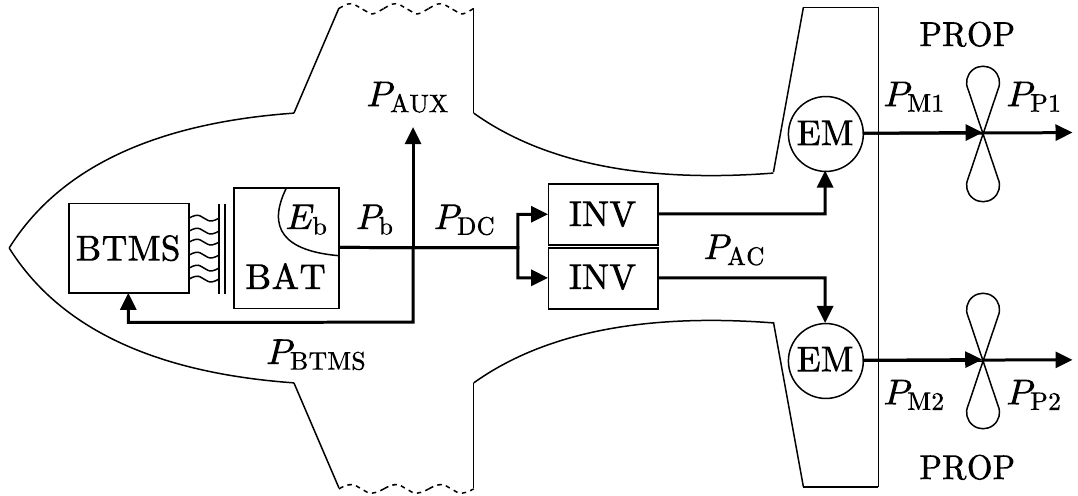}
    \caption{Power block diagram showing the electric aircraft powertrain topology. It consists of a BTMS, a battery system (BAT), two inverters (INV), two electric motors (EM) and two propellers (PROP). The power flows (indicated by arrows), $P_i$, are battery output power, BTMS, inverter, and machine input power flows indicated with the subscripts for $i =$ \{b, BTMS, DC, AC\}. Others are auxiliary loads, $P_\mathrm{aux}$, and in-/output power to the propellors, $P_\mathrm{M}$ and $P_\mathrm{P}$, respectively. The placement of some components in this schematic does not correspond to their actual location in the aircraft and is for visual convenience only.}
    \label{fig:plane}
    \vspace{-5pt}
\end{figure}

\textit{Related literature:} This study investigates two primary research areas. The first focuses on modeling electric vehicle power consumption, including battery system electrical and thermal modeling, and addressing thermal runaway behavior. Prior works have modeled ground vehicle battery systems at cell or system levels \cite{Reiter2019ASystems,Wilkins2017ModelPack} but lacked thermal conditioning to prevent overheating \cite{Schomann2012Hybrid-ElectricAircraft}. Modeling efforts for fully electric and hybrid-electric aircraft battery systems (BTMS) and thermal management have been limited. While some research emphasizes precise battery thermal estimation for mission planning \cite{Chin2019BatteryX-57}, connections among individual cells are often overlooked, which limits the model fidelity. Other studies explore cooling technologies for hybrid-electric aircraft \cite{Kabir2018InvestigationAirplane,Rheaume2019CommercialImprovements}, investigating factors influencing cooling performance and thermal runaway risk. Unlike hybrid-electric aircraft, where the battery system can be shut off during unsafe operation, fully electric aircraft lack this option, risking complete power loss. The second research area focuses on minimum-energy design for battery system thermal conditioning, using BTMS. Research in electronics often relies on forced air cooling at the module level, optimizing module spacing for efficient cooling \cite{Kirad2021DesignSystem}. Aviation-specific studies, such as for a 19-seater single-aisle hybrid-electric aircraft, have optimized BTMS design, assessing various heat pump technologies \cite{Kellermann2022DesignAircraft}. However, research for fully electric aircraft is lacking. Currently, no comprehensive studies address system design challenges for electric aviation at the battery system level, considering the impact of BTMS on flight range and energy consumption. Additionally, there is limited exploration of thermal runaway behavior in worst-case scenarios.

\textit{Statement of contributions:} This work presents a mathematical framework that simulates the electrical power consumption of a compact 8-seater electric aircraft, with powertrain architecture shown in Fig.~\ref{fig:plane}, across a predefined flight regime. The aircraft is equipped with high-specific energy density nickel manganese cobalt (NMC) Lithium-ion cells. Subsequently, it determines the heat generation of a pouch-type NMC battery system at cell level \cite{Stefan2020HighTechnology} to investigate if thermal management in the form of BTMS is needed. The framework then facilitates BTMS design optimization with a minimal-energy objective to keep the battery system within safe operation limits. Additionally, we define the thermal behavior of the cell under heating-induced thermal runaway (TR). Finally, we showcase our framework on a battery system design for an 8-seater fully electric lightweight aircraft, by determining theoretical flight range without BTMS and subsequently incorporating BTMS and minimizing the battery energy consumed over the flight regime.

\textit{Organization:} Section \ref{sec:methodology} presents the powertrain component modeling methods, the battery modeling approach and the BTMS modeling method for the aircraft design. It additionally introduces the thermal runaway modeling method and the minimum-energy-consumption design problem. Lastly, we discuss some limitations of this framework. In Section \ref{sec:results}, we showcase the framework for a fixed flight regime. Conclusions are drawn in Section \ref{sec:conclusion}, after which we provide an outlook on future research.

%% file: Sections/Methodology.tex
\section{Methodology}\label{sec:methodology}
This section presents the design optimization problem and its constraints that describe the electric aircraft powertrain, battery system and BTMS. We introduce the objective and aircraft powertrain component models, whereby we put particular emphasis on the battery model. Next, we extract the battery parameters from test data  and introduce the thermal model for the battery system and BTMS. Subsequently, we introduce the battery thermal runaway model and summarize the problem. The aircraft powertrain we consider in this paper is shown in Fig. \ref{fig:plane}.

\subsection{Objective}\label{sec:objective}
The objective in our optimization problem is to minimize
the internal energy consumption of the battery over a fixed flight regime starting at $t$ = 0 to the end time $t = t_\mathrm{f}$, where $t_\mathrm{f}$ is the final time of the cycle. We define the internal battery energy consumption as
\par\nobreak
\begingroup
\allowdisplaybreaks
\begin{small}
\begin{equation}\label{eq:minEb}
    \min_{p}~\Delta E_\mathrm{b}(p),
\end{equation}
\end{small}%
\endgroup
where $\Delta E_\mathrm{b}(p)$ is equal to the difference in internal battery energy as a function of $p$, which represents the set of design variables, $p=(T_\mathrm{fl}, \dot{V}_\mathrm{fl}, P_\mathrm{BTMS,rated}$), being the temperature of the cooling fluid, the volumetric flow rate of the cooling fluid and the rated power of the BTMS, respectively. The state variable in this problem is battery energy $x(t) = E_\mathrm{b}$. The difference of internal battery energy is defined by
\par\nobreak
\begingroup
\allowdisplaybreaks
\begin{small}
\begin{equation}\label{eq:deltaEb}
    \Delta E_\mathrm{b}(p) = E_\mathrm{b}(0)-E_\mathrm{b}(t_\mathrm{f}),
\end{equation}
\end{small}%
\endgroup
where $E_\mathrm{b}(0)$ and $E_\mathrm{b}(t_\mathrm{f})$ are the internal battery energy at the beginning and at the end of the flight regime, respectively.

\subsection{Longitudinal Aircraft Dynamics}\label{sec:dynamics}
In this section, we model the aircraft as a longitudinal point mass with the required thrust output as a load on the electric motors using a quasi-static backward-facing modeling approach  
in the time domain. We consider a provided flight profile consisting of an exogenous speed trajectory $v(t)$ and height trajectory $h(t)$. Assuming that the pilot takes the actions necessary to follow the mission and that the mass of the aircraft remains constant over the flight cycle, the required thrust provided by the propellers is expressed by longitudinal point mass equation of motion as
\par
\begin{figure}[t]
    \centering
    \includegraphics[width=0.85\linewidth]{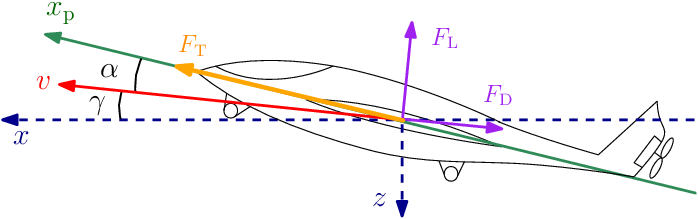}
    \caption{Applied coordinate system for aircraft in flight for the use of a longitudinal point mass model, with components $F_\mathrm{L}$ and $F_\mathrm{D}$, respectively, representing total lift and drag force. The flight path angle $\gamma$ represents the direction of the flight, while the angle of attack $\alpha$ represents the angle of the wings w.r.t.\ the incoming air.  }
    \label{fig:Planecoord}
    \vspace{-10pt}
\end{figure}

\nobreak
\begingroup
\allowdisplaybreaks
\begin{small}
\begin{equation}\label{eq:PointMass}
    F_\mathrm{T}(t) = \frac{\frac{1}{2}\cdot \rho(h(t))\cdot v^2(t)\cdot C_\mathrm{d}\cdot S_\mathrm{w}+m\cdot (g\cdot \sin(\gamma)+\dot{v}(t))}{\cos(\alpha)},
\end{equation}
\end{small}%
\endgroup
where $\rho(h(t))$ is the air density as a function of aircraft height $h(t)$, $v(t)$ is the airspeed of the aircraft, $C_\mathrm{d}$ is the drag coefficient, $S_\mathrm{w}$ is the wing area, $m$ is the mass of the aircraft $g$ is Earth's gravitational constant, $\gamma$ is the flight path angle and $\alpha$ is the angle of attack \cite{Cook2007FlightPrinciples}. The applied coordinate system for the longitudinal aircraft dynamics is shown in Fig.~\ref{fig:Planecoord}. In this study, we define the total mass $m$ of the aircraft as
\par\nobreak
\begingroup
\allowdisplaybreaks
\begin{small}
\begin{equation}\label{eq:totalmass}
    m = m_\mathrm{empty}+m_\mathrm{b}+m_\mathrm{BTMS},
\end{equation}
\end{small}%
\endgroup
where $m_\mathrm{empty}$ is the total mass of the plane without the mass of the battery system, $m_\mathrm{b}$, and mass of BTMS, $m_\mathrm{BTMS}$.

\subsection{Powertrain model}\label{sec:powertrain}
In this study, we assume that both aircraft motors and both inverters perform identically over the flight cycle, thereby lumping them together to a single unit. For the motors, we consider axial flux permanent magnet synchronous machines (PMSM). To obtain the required electric motor power over the flight regime, we model a constant rotational speed, variable-pitch propeller connected to the motor shaft of both electric motors as a unified unit with
\par\nobreak
\begingroup
\allowdisplaybreaks
\begin{small}
\begin{equation}\label{eq:motorpower}
    P_\mathrm{m}(t) = \frac{F_\mathrm{T}(t)\cdot{v(t)}}{\eta_\mathrm{p}},
\end{equation}
\end{small}%
\endgroup
where $\eta_\mathrm{p}$ is the propulsion efficiency of the propeller which accounts for the viscous profile drag on the blades, and for the kinetic energy lost in the accelerated airflow. Accordingly, the required motor torque is defined by
\par\nobreak
\begingroup
\allowdisplaybreaks
\begin{small}
\begin{equation}\label{eq:torque}
    \tau_\mathrm{m}(t) = \frac{P_\mathrm{m}(t)}{\omega_\mathrm{m}},
\end{equation}
\end{small}%
\endgroup
where $\omega_\mathrm{m}=\omega_\mathrm{p,cons}$ is the rotational speed of the motor, which is equal to the propellor speed $\omega_\mathrm{p,cons}$. From a modeling standpoint, the motor power is only limited by the maximum torque $\tau_\mathrm{m,max}$ since the rotational speed is constant. Regeneration is not considered. 
The electrical power defined by the alternating current (AC) demanded at the inverter output is labeled $P_\mathrm{AC}(t)$ and is expressed as
\par\nobreak
\begingroup
\allowdisplaybreaks
\begin{small}
\begin{equation}\label{eq:ACpower}
    P_\mathrm{AC}(t) = \frac{1}{\eta_\mathrm{m}(\tau_\mathrm{m}(t))}\cdot P_\mathrm{m}(t),
\end{equation}
\end{small}%
\endgroup
where $\eta_\mathrm{m}$ is the motor efficiency as a function of motor torque for a constant rotational speed. The inverter provides power to the motor by rapidly alternating the direction of a direct current (DC) input, converting it into AC. Throughout this operation, there are experienced losses due to the rapid switching. We derive a quadratic model for these inverter losses, applying a general quadratic power loss model of the form
\par\nobreak
\begingroup
\allowdisplaybreaks
\begin{small}
\begin{equation}\label{eq:DCpower}
    P_\mathrm{DC}(t) = \beta \cdot P_\mathrm{AC}^2(t)+P_\mathrm{AC}(t),
\end{equation}
\end{small}%
\endgroup
where $\beta$ is an efficiency parameter, subject to identification.


\subsection{Battery model}\label{sec:Battery}
The power at the battery system terminals is equal to the sum of all power consumers in the aircraft, defined by
\par\nobreak
\begingroup
\allowdisplaybreaks
\begin{small}
\begin{equation}\label{eq:Pb}
    P_\mathrm{b}(t) = P_\mathrm{DC}(t)+P_\mathrm{BTMS}(t)+P_\mathrm{aux},
\end{equation}
\end{small}%
\endgroup
where $P_\mathrm{BTMS}(t)$ represents the power drawn by the BTMS, and $P_\mathrm{aux}$ models a constant auxiliary power flow. In this context, the internal battery energy, $E_\mathrm{b}$, changes with demanded internal battery power $P_\mathrm{i}$ as
\par\nobreak
\begingroup
\allowdisplaybreaks
\begin{small}
\begin{equation}\label{eq:dotEb}
    \dot{E}_\mathrm{b}(t) = -P_\mathrm{i}(t),
\end{equation}
\end{small}%
\endgroup
where the minus sign ensures the battery is discharged when $P_\mathrm{b}(t)$ is positive. Additionally, $E_\mathrm{b}(t)$ is bounded by
\par\nobreak
\begingroup
\allowdisplaybreaks
\begin{small}
\begin{equation}\label{eq:EbRange}
    E_{\mathrm{b}}(t) \in\left[E_{\mathrm{b,min}}, E_{\mathrm{b,max}}\right],
\end{equation}
\end{small}%
\endgroup
where $E_{\mathrm{b,max}}$ and $E_{\mathrm{b,min}}$ correspond with the minimum and maximum battery state of charge levels, respectively, $\zeta_{\mathrm{b,min}}$ and $\zeta_{\mathrm{b,max}}$ by establishing the correlation between the variables in a lookup table. We assume the aircraft starts the flight regime with a fully charged battery
\par\nobreak
\begingroup
\allowdisplaybreaks
\begin{small}
\begin{equation}\label{eq:Eb0}
    E_{\mathrm{b}}(0)=\zeta_{\mathrm{b,max}}\cdot E_{\mathrm{b}, \max }.
\end{equation}
\end{small}%
\endgroup
The internal battery power is defined as 
\par\nobreak
\begingroup
\allowdisplaybreaks
\begin{small}
\begin{equation}\label{eq:P_i}
    P_\mathrm{i}(t) = U_\mathrm{OC}(\zeta(t))\cdot I_\mathrm{b}(t),
\end{equation}
\end{small}%
\endgroup
where $U_\mathrm{OC}$ is the open circuit voltage (OCV), defined by the cell's state of charge $\zeta(t)$.
To obtain the battery current draw from the power demand, we apply
\par
\begin{figure}[t]
    \centering
    \includegraphics[width=0.8\linewidth]{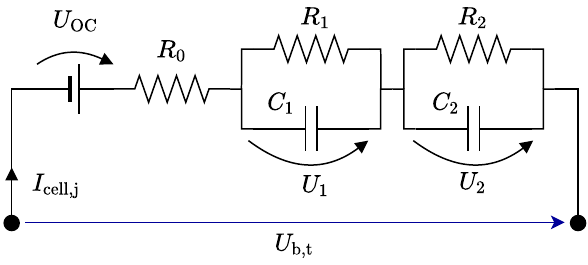}
    \caption{Second-order equivalent circuit model diagram, consisting of an open-circuit voltage source $U_\mathrm{oc}$, internal resistance $R_\mathrm{0}$ and two RC pairs.}
    \label{fig:2RC}
    \vspace{-10pt}
\end{figure}

\nobreak
\begingroup
\allowdisplaybreaks
\begin{small}
\begin{equation}\label{eq:curr_draw}
    I_\mathrm{b}(t) = \frac{P_\mathrm{b}(t)}{n_\mathrm{s}\cdot U_\mathrm{cell,t}(t)},
\end{equation}
\end{small}%
\endgroup
where $n_\mathrm{s}$ is the number of cells in series and $U_\mathrm{cell,t}(t)$ is the battery cell's terminal voltage, which we obtain by applying an equivalent circuit model (ECM) with two RC-branches to provide a good trade-off between complexity and accuracy\cite{Hu2012ABatteries}.
Fig. \ref{fig:2RC} illustrates this model, whereby component values vary with temperature and state of charge (SOC) \cite{Wilkins2017ModelPack}.
The battery cell studied is an 11.84 Ah high specific energy density (HED) cell employing nickel manganese cobalt (NMC) chemistry, renowned for its balance of energy, power, cycle life, and thermal stability \cite{Blomgren2017TheBatteries}.

Next, we outline a single-cell modeling approach by a parallel circuit with $n_\mathrm{p}$ cells, employing an ECM featuring two RC elements. When $n_\mathrm{p}$ equals 1, the parallel circuit model transforms into that of a single cell. Therefore, by applying the total current draw defined in \eqref{eq:curr_draw}, we model the current draw $I_\mathrm{cell,j}(t)$ of every cell  $j \in \mathbb{N}\,\cap\, [1,n_\mathrm{p}]$ in the parallel circuit as
\par\nobreak
\begingroup
\allowdisplaybreaks
\begin{small}
\begin{equation}\label{eq:Icell}
 I_\mathrm{cell,j}(t) = \frac{I_\mathrm{b}(t)}{n_\mathrm{p}}.
\end{equation}
\end{small}%
\endgroup


Following the second-order ECM shown in Fig.~\ref{fig:2RC}, the terminal voltage of the cell under load is defined as
\par\nobreak
\begingroup
\allowdisplaybreaks
\begin{small}
\begin{equation}\label{eq:Ucell}
    U_\mathrm{cell,j}(t) = U_\mathrm{OC}(\zeta(t))-I_\mathrm{cell,j}(t)\cdot R_0-U_\mathrm{1}(t)-U_\mathrm{2}(t),
\end{equation}
\end{small}%
\endgroup
where $R_0$ is the cell's internal series resistance, $U_n(t)$ is the over-voltage over the $n$-th RC element, here with $n \in \{1, 2\}$.


\begin{figure}[t]
    \centering
    \includegraphics[width=0.85\linewidth]{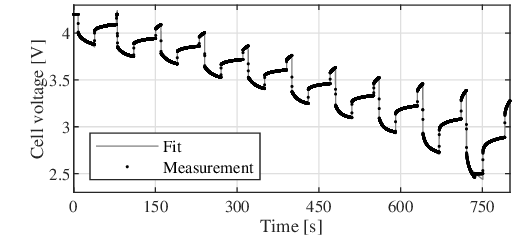}
    \caption{Fit of HPPC cycle voltage for the 11.84 Ah NMC cell at 23 $^\circ \mathrm{C}$. The RMSE of the fit is 10.08 mV (0.58\%).}
    \label{fig:HPPCfit}
    \vspace{-10pt}
\end{figure}

We utilize the voltage and current data collected from the HPPC tests to fit the Equivalent Circuit Model (ECM) parameters with nonlinear regression utilizing a least-squares approach \cite{Tran2021ComparativeNCA}. 
The result of HPPC data fit is shown in Fig.~\ref{fig:HPPCfit}. 

\subsection{Battery thermal model}\label{sec:thermal}
To minimize the workload in modeling and testing, LIBs are often treated as lumped thermal masses with consistent heat capacity. 
We assume uniform or piecewise uniform temperature distribution within the battery, enabling the feasibility of employing lumped elements for characterizing the battery's thermal dynamics. 
We can simplify the energy balance of every cell  $j \in \mathbb{N}\,\cap\, [1,n_\mathrm{p}]$ in the parallel circuit as
\par\nobreak
\begingroup
\allowdisplaybreaks
\begin{small}
\begin{equation}\label{eq:enbal}
    c_\mathrm{p,j}\cdot m_\mathrm{cell,j}\cdot\dot{T}_\mathrm{cell,j}(t) = \dot{Q}_\mathrm{gen,j}(t)+\dot{Q}_\mathrm{diss,j}(t),
\end{equation}
\end{small}%
\endgroup
where $c_\mathrm{p,j}$ is the specific heat capacity of the cell, $m_\mathrm{cell,j}$ is the mass of the cell, $T_\mathrm{cell,j}(t)$ is the cell's temperature, $\dot{Q}_\mathrm{gen,j}(t)$ is the cell's heat generation rate and $\dot{Q}_\mathrm{diss,j}(t)$ is the heat dissipation rate. We consider both convective $\dot{Q}_{\mathrm{conv,j}}(t)$ and radiative heat rate $\dot{Q}_{\mathrm{rad,j}}(t)$, defined by
\par\nobreak
\begingroup
\allowdisplaybreaks
\begin{small}
\begin{equation}\label{eq:dQdiss}
    \dot{Q}_{\mathrm{diss,j}}(t)= \dot{Q}_{\mathrm{conv,j}}(t)+\dot{Q}_{\mathrm{rad,j}}(t),
\end{equation}
\end{small}%
\endgroup
and
\par\nobreak
\begingroup
\allowdisplaybreaks
\begin{small}
\begin{equation}\label{eq:dQconv}
    \dot{Q}_{\mathrm{conv,j}}(t) = A\cdot h\cdot\left(T_{\mathrm{cell,j}}(t)-T_{\infty,j}(t)\right),
\end{equation}
\end{small}%
\endgroup
where $A$ is the cell's surface, $h$ is the heat transfer coefficient, and $T_{\infty}$ is the environmental temperature around the cell, and with radiative heat rate defined by
\par\nobreak
\begingroup
\allowdisplaybreaks
\begin{small}
\begin{equation}\label{eq:dQrad}
    \dot{Q}_{\mathrm{rad,j}}(t) = A\cdot\varepsilon\cdot\sigma\cdot\left(T_{\mathrm{cell,j}}^4(t)-T_{\infty,j}^4(t)\right),
\end{equation}
\end{small}%
\endgroup
where $\varepsilon$ is the surface emissivity of the cell, taken as $\varepsilon$ = 0.8 for this application 
and $\sigma$ is the Stefan-Boltzmann constant ($\sigma$ = 5.67$\cdot$10$^{-8}$ W$\cdot$m$^{-2}\cdot$K$^{-4}$).
To model the heat generation rate $\dot{Q}_\mathrm{gen,j}(t)$ of a battery on cell level, we apply an electric‐thermal coupling model based on the assumption of uniform heat generation inside the battery cell proposed by Bernardi et al. \cite{Bernardi1984GeneralSystems}. The coupling model consists of two heat sources, $\dot{Q}_\mathrm{irr,j}(t)$, the irreversible heat generated by the internal resistance of the cell and $\dot{Q}_\mathrm{rev,j}(t)$, the reversible heat generation because of the entropy change due to the chemical reactions inside the cell. The heat generation rate can, therefore, be defined by
\par\nobreak
\begingroup
\allowdisplaybreaks
\begin{small}
\begin{equation}\label{eq:dQgen}
    \dot{Q}_\mathrm{gen,j}(t) = \dot{Q}_\mathrm{irr,j}(t)+\dot{Q}_\mathrm{rev,j}(t),
\end{equation}
\end{small}%
\endgroup
with
\par\nobreak
\begingroup
\allowdisplaybreaks
\begin{small}
\begin{equation}\label{eq:dQirr}
    \dot{Q}_\mathrm{irr,j}(t) = I_\mathrm{cell,j}(t)\cdot ( U_\mathrm{OC}(\zeta(t))-U_\mathrm{cell,j}(t)),
\end{equation}
\end{small}%
\endgroup
and
\par\nobreak
\begingroup
\allowdisplaybreaks
\begin{small}
\begin{equation}\label{eq:dQrev}
    \dot{Q}_\mathrm{rev,j}(t) = I_\mathrm{cell,j}(t)\cdot T_\mathrm{cell,j}(t)\cdot \frac{\partial U_\mathrm{OC}}{\partial T_\mathrm{cell,j}},
\end{equation}
\end{small}%
\endgroup
where $\frac{\partial U_\mathrm{OC}}{\partial T_\mathrm{cell,j}}$ is the entropy coefficient of the cell as introduced in \cite{Eddahech2013ThermalChanges}. For the cell-level simulation of a battery system comprising $n$ cells, we omit consideration of thermal radiation as defined in \eqref{eq:dQrad}. Consequently, temperatures may experience a slight overestimation, adding an inherent margin to the model. 


\subsection{Battery Thermal Management System model}\label{sec:BTMS}

\begin{figure}[t]
    \centering
    \includegraphics[width=0.8\linewidth]{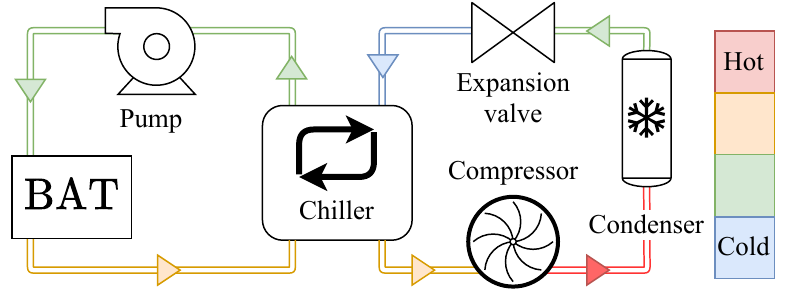}
    \caption{BTMS system layout. Battery coolant loop left of the chiller, VCM on the right side. The actively controlled components are the pump and the compressor. The chiller represents a heat exchanger between the two loops.}
    \label{fig:BTMS}
    \vspace{-10pt}
\end{figure}

A Battery Thermal Management System (BTMS) can be passive, using heat pipes and phase-change materials, or active, employing forced-air or liquid cooling systems \cite{Kellermann2022DesignAircraft}. Active systems utilize a liquid coolant to transfer heat from the battery to dissipation. Their effectiveness depends on factors such as fluid selection, mass flow, temperature control, and heat sink characteristics. In this study, we model an active BTMS incorporating a Vapor Cycle Machine (VCM), an established type of refrigeration system where a refrigerant undergoes phase changes to remove heat. The BTMS, comprising VCM, chiller, and coolant loop, is depicted in Fig. \ref{fig:BTMS}.
We conceptualize the BTMS as a closed loop, where a fluid circulates through cooling channels integrated into aluminium cold plates. These cold plates function as heat sinks and are in direct physical contact with the battery cells. One of the main advantages of this BTMS layout is the high coefficient of performance, $K_\mathrm{\textbf{BTMS}}$, depicting the ratio of useful heat energy produced to electrical energy consumption. The electrical power consumption of the BTMS, $P_\mathrm{BTMS}$, influences the total electrical power consumption as stated in (\ref{eq:Pb}), which is modeled as
\par\nobreak
\begingroup
\allowdisplaybreaks
\begin{small}
\begin{equation}\label{eq:PBTMS}
    P_\mathrm{BTMS}(t) = \frac{\dot{Q}_\mathrm{BTMS}(t)}{K_\mathrm{BTMS}},
\end{equation}
\end{small}%
\endgroup
where $\dot{Q}_\mathrm{BTMS}(t)$ is the heat dissipated by the fluid. 
As the mass and the electrical energy consumption of the BTMS both contribute to the total energy consumption of the aircraft, we express the mass of the BTMS as
\par\nobreak
\begingroup
\allowdisplaybreaks
\begin{small}
\begin{equation}\label{eq:mBTMS}
    m_\mathrm{BTMS} = m_\mathrm{loop}+m_\mathrm{VCM},
\end{equation}
\end{small}%
\endgroup
where $m_\mathrm{loop}$ and $m_\mathrm{VCM}$ represent the mass of the VCM and the mass of the coolant loop, respectively. The mass of the VCM is scaled with cooling power density $\rho_\mathrm{P}$ and defined as
\par\nobreak
\begingroup
\allowdisplaybreaks
\begin{small}
\begin{equation}\label{eq:mVCM}
    m_\mathrm{VCM} = \frac{P_\mathrm{BTMS,rated}}{\rho_\mathrm{P}},
\end{equation}
\end{small}%
\endgroup
where $P_\mathrm{BTMS,rated}$ is the rated cooling power of the BTMS and $\rho_\mathrm{P}$ is the specific power density of the VCM, expressed in cooling power per unit of mass. For electrical power, we need to divide this value by the coefficient of performance, $K_\mathrm{BTMS}$. 

The addition of a BTMS enables the system to dissipate heat via a cooling medium, thereby changing the dissipative properties of the environment around the cell, compared to purely natural heat dissipation. Therefore, we extend \eqref{eq:dQdiss} to 
\par\nobreak
\begingroup
\allowdisplaybreaks
\begin{small}
\begin{equation}\label{eq:dQdiss2}
    \dot{Q}_{\mathrm{diss,j}}(t)= \dot{Q}_{\mathrm{conv,j}}(t)+\dot{Q}_{\mathrm{rad,j}}(t)+\dot{Q}_{\mathrm{BTMS,j}}(t),
\end{equation}
\end{small}%
\endgroup
with
\par\nobreak
\begingroup
\allowdisplaybreaks
\begin{small}
\begin{equation}\label{eq:dQBTMS}
    \dot{Q}_{\mathrm{BTMS,j}}(t)= A_\mathrm{BTMS,j}\cdot h_\mathrm{BTMS} \cdot\left(T_\mathrm{cell,j}(t)-T_{\mathrm{fl,j}}(t)\right),
\end{equation}
\end{small}%
\endgroup
where $A_\mathrm{BTMS,j}$ represents the heat transferring area of the BTMS at cell $j$, $h_\mathrm{BTMS}$ denotes the mean heat transfer coefficient of the BTMS, and $T_{\mathrm{fl,j}}$ signifies the inlet fluid temperature. We assume an isobaric, incompressible, and frictionless fluid within the BTMS. Flow through the BTMS is considered one-dimensional and strictly laminar.
We employ the correlation for tube banks \cite{Incropera2011FundamentalsEdition}, ensuring conditions for cross- and longitudinal pitch ratios ($a <$ 1.2, $b/a <$ 1.0, and $\mathrm{Re}<$10$^4$). Here, $a$ represents the transverse pitch ratio ($s_1/D_\mathrm{h}$), $b$ is the longitudinal pitch ratio ($s_2/D_\mathrm{h}$), $s_t$ is the transverse pitch, and $s_l$ is the longitudinal pitch between channel centers. The channel, modeled as a plate duct, is formed by placing two plates at a fixed distance $s_\mathrm{ch}$ from each other. 
The mean convective heat transfer coefficient for the BTMS applied in \eqref{eq:dQBTMS} can be defined as
\par\nobreak
\begingroup
\allowdisplaybreaks
\begin{small}
\begin{equation}\label{eq:hBTMS}
 h_\mathrm{BTMS} = \frac{k_\mathrm{fl}\cdot\overline{\mathrm{Nu}}}{D_\mathrm{h}},
\end{equation}
\end{small}%
\endgroup
where ${k_\mathrm{fl}}$ is the thermal conductivity of the fluid, $D_\mathrm{h}=2\cdot s_\mathrm{ch}$ the hydraulic diameter and $\overline{\mathrm{Nu}}$ is the mean Nusselt number. We apply approximate expressions defined for parallel plate ducts defined by Gnielinski \cite{Gnielinski2010G2Ducts} to reduce the complexity of the simulation. 
For strictly laminar flow, the mean Nusselt number in parallel plate ducts is given by
\par\nobreak
\begingroup
\allowdisplaybreaks
\begin{small}
\begin{equation}\label{eq:meanNusselt}
\overline{\mathrm{Nu}}=\sqrt[3]{\mathrm{Nu}_1^3+\mathrm{Nu}_2^3},
\end{equation}
\end{small}%
\endgroup
where $\mathrm{Nu}_1$ = 4.816 for one heat-exchanging wall and 
\par\nobreak
\begingroup
\allowdisplaybreaks
\begin{small}
\begin{equation}\label{eq:Nusselt2}
    \mathrm{Nu}_2 = 1.841\cdot \sqrt[3]{\mathrm{Re}\cdot \mathrm{Pr}\cdot\frac{D_\mathrm{h}}{l}},
\end{equation}
\end{small}%
\endgroup
where $\operatorname{Re}$ is the Reynolds number, $\operatorname{Pr}$ is the Prandtl number ($\operatorname{Pr}>0.6$) \cite{Incropera2011FundamentalsEdition} and $l$ is the length of the cooling channel. 
The Reynolds number is equal to
\par\nobreak
\begingroup
\allowdisplaybreaks
\begin{small}
\begin{equation}\label{eq:Reynolds}
    \mathrm{Re} = \frac{ D_\mathrm{h}\cdot v_\mathrm{fl}}{\nu_\mathrm{fl}},
\end{equation}
\end{small}%
\endgroup
where $v_\mathrm{fl}$ is the velocity of the fluid, $\rho_\mathrm{fl}$ is the density of the fluid and $\nu_\mathrm{fl}$ is the kinematic viscosity of the fluid. The velocity of the fluid is expressed as
\par\nobreak
\begingroup
\allowdisplaybreaks
\begin{small}
\begin{equation}\label{eq:vfl}
    v_\mathrm{fl} = \frac{\dot{V}_\mathrm{fl}}{A_\mathrm{cross,ch}},
\end{equation}
\end{small}%
\endgroup
where $\dot{V}_\mathrm{fl}$ is the volumetric flow rate of the cooling fluid and $A_\mathrm{cross,ch}$ is the area of a channel's cross-section. The volumetric flow rate of the cooling fluid can be adjusted between zero and the maximum allowable flow rate for a certain channel size.
As the fluid passes the cells within a cooling channel, it is heated up. Consequently, cells located at the outlet of the channel experience less cooling compared to those at the inlet. 
\subsection{Thermal runaway model}\label{sec:TR}

Understanding and predicting Thermal Runaway (TR) mechanisms is crucial for establishing early warning triggers in battery thermal management and devising effective prevention strategies. 
In this section, we adopt a simplified approach for modeling decomposition reactions proposed by \cite{Lalinde2021AppliedBatteries}, utilizing measurements from non-destructive accelerating rate calorimeter (ARC) tests. All decomposition reactions are grouped into a single global reaction for the cell \cite{Lalinde2021AppliedBatteries}, while Equations \eqref{eq:enbal}-\eqref{eq:dQrad} are retained since the heat dissipation rate $\dot{Q}_\mathrm{diss}(t)$ is independent of decomposition reactions. The generated heat flow $\dot{Q}_\mathrm{gen}(t)$ describes the chemical reactions within the cell. For the thermal runaway scenario, we therefore redefine generated heat as
\par\nobreak
\begingroup
\allowdisplaybreaks
\begin{small}
\begin{equation}\label{eq:dQgenTR}
     \dot{Q}_\mathrm{gen,j}(t) = m_\mathrm{cell}\cdot i_\mathrm{react} \cdot \dot{x}_\mathrm{cell}(t),
\end{equation}
\end{small}%
\endgroup
where $i_\mathrm{react}$ is the specific enthalpy of the cell reaction and $\dot{x}_\mathrm{cell}(t)$ is the conversion reaction rate, expressed as
\par\nobreak
\begingroup
\allowdisplaybreaks
\begin{small}
\begin{equation}\label{eq:dXTR}
    \dot{x}_\mathrm{cell}(t) = A_\mathrm{x}\cdot x_\mathrm{cell}\cdot \exp{\left(\frac{-E_\mathrm{A}}{k_\mathrm{b}\cdot T_\mathrm{cell}}\right)},
\end{equation}
\end{small}%
\endgroup
where $A_\mathrm{x}$ is the frequency factor, $x_\mathrm{cell}$ is the degree of conversion with $x_\mathrm{cell}\in[0,1]$, $E_\mathrm{A}$ is the activation energy of the reaction, and $k_\mathrm{b}$ is the Boltzmann constant ($k_\mathrm{b}$ = 1.38$\cdot10^{-23}$ J$\cdot$K$^{-1}$). We initiate the reaction at $x_\mathrm{cell}$ = 1, indicating that no reactant is consumed \cite{Melcher2016ModelingParameters}. For the NMC type LIB applied in this study, the three values that characterize the overall chemical reaction of the cell, $i_\mathrm{react}$, $A_\mathrm{x}$ and $E_\mathrm{A}$, can be estimated from the cell temperature and temperature rate\cite{Ohneseit2023ThermalARC}. Since the validation of this model against real-world data requires destructive tests, we deem this validation to be beyond the scope of the current paper and devote it to future research.
\subsection{Optimization problem}\label{sec:Problem}
The objective is to minimize the battery energy consumption of the electric aircraft over a predefined flight regime. The capacity  of its battery system $E_\mathrm{b,max}=304$kWh is fixed as well, due to volume requirements. Therefore, we minimize the battery by finding the optimal BTMS size to retain sufficient cooling capacity and to avoid unnecessary weight increase. For this problem, the state variable are battery energy and cell temperature $x(t) = (E_\mathrm{b},T_\mathrm{cell})$. The design variables are $p=(T_\mathrm{fl}, \dot{V}_\mathrm{fl}, P_\mathrm{BTMS,rated}$), being the temperature of the cooling fluid, the volumetric flow rate of the cooling fluid and the rated power of the BTMS, respectively.
\begin{prob}[Nonlinear design problem]\label{prob:main}
The minimum energy design is the solution of
\begin{equation*}
\begin{aligned}
&\!\min_{p} & &\Delta E_\mathrm{b}(p), \\
& \textnormal{s.t. }  & & P_\mathrm{b,min}\leq P_\mathrm{b}(t,p) \leq P_\mathrm{b,max},\\
& & &U_\mathrm{cell, min}\leq U_\mathrm{cell}(t,p) \leq U_\mathrm{cell,max},\\
& & &T_\mathrm{cell,opt,min}\leq T_\mathrm{cell}(t,p) \leq T_\mathrm{cell,opt,max},\\
& & &E_{\mathrm{b}}(t) \in\left[\zeta_{\mathrm{b,min}}, \zeta_{\mathrm{b,max}}\right] \cdot E_{\mathrm{b}, \max },\\
& & & 0\leq I_\mathrm{b}(t,p)\leq I_\mathrm{b,max},\\
& & & P_\mathrm{BTMS}(t,p)\leq P_\mathrm{BTMS,rated},\\
& & &\dot{V}_\mathrm{fl,min}\leq \dot{V}_\mathrm{fl}(t) \leq \dot{V}_\mathrm{fl,max},\\
& & & \eqref{eq:deltaEb}-\eqref{eq:Ucell}, \eqref{eq:enbal}-\eqref{eq:dQconv}, \eqref{eq:dQgen}-\eqref{eq:dQrev}, \eqref{eq:PBTMS}-\eqref{eq:dQBTMS}.
\end{aligned}
\end{equation*}
\end{prob}

%% file: Sections/Results.tex
\section{Results}\label{sec:results}

In this section, we present the battery system design as a result of the design constraints and requirements for the 8-seater light electric aircraft. Next, we present the numerical results of the optimal BTMS design. Lastly, we observe the TR behavior of the applied NMC cell. 

\subsection{Battery system design}
\begin{figure}[t]
    \centering
    \includegraphics[width=0.85\linewidth]{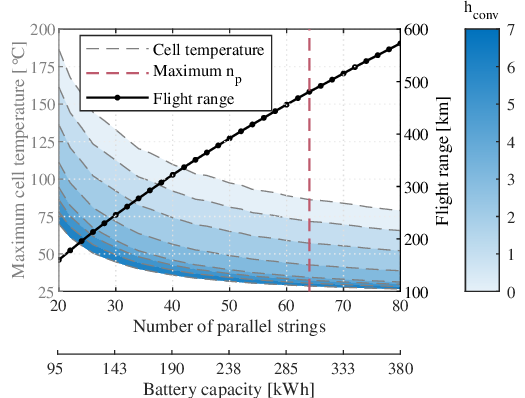}
    \caption{Flight range and maximum cell temperature over the number of parallel battery strings, for varying heat transfer coefficient values $h$ from 0 of 7 W/m$^2$K. Steady-state flight at 500 m from 100-15\% SOC and without BTMS added. Maximum parallel switched cell strings is 64 (red dotted line).}
    \label{fig:range}
\end{figure}

Considering the constraints and framework outlined in Section \ref{sec:methodology}, we propose a baseline battery design for an 8-seater electric aircraft with a \unit[304]{ kWh} battery system, including BTMS. The design ensures failure redundancy by integrating two parallel individually switched battery banks, each at \unit[400]{V}. These banks, connected to separate pairs of inverters and EMs, can be centrally controlled. 
The system features a 118-64 series-parallel cell layout to fit the aircraft's volume requirement. 
We analyze the theoretical flight range for an electric aircraft without BTMS, considering steady-state flight at 500 m and 180 km/h, discharging the battery from 100 to 15\% SOC. The total system voltage constraint of 400 V requires the number of cells in series to remain constant. However, the number of parallel strings can vary. Fig. \ref{fig:range} illustrates this simulation across increasing parallel strings, affecting battery capacity and aircraft weight. The simulation considers multiple heat convection coefficients ($h$ = 0 to $h$ = 7), from no heat transfer to average convective heat transfer. The theoretical approach without convective heat transfer is impractical, with real-world scenarios always involving some form of heat transfer. Furthermore, it is observed that increasing parallel cells correlates with lower maximum cell temperatures. At $h$ = 7, the temperature remains within the optimal range (\unit[28.5]{$^\circ$C}). However, variations in heat convection levels throughout the flight cycle may push cells beyond this range. For instance, at $h$ = 3.5, the temperature exceeds the optimal range (\unit[35.4]{$^\circ$C}), jeopardizing operational safety. Additionally, increasing battery capacity could offer advantages, evident in the range curve's slight deflection tendency (Fig. \ref{fig:range}).

\begin{table}[t]
\centering
\caption{Numerical results for the minimum energy BTMS design. The theoretical range is calculated purely based on increased power consumption of the BTMS, disregarding temperature limits. }
\label{tab:Results}
\resizebox{\columnwidth}{!}{%
\def\arraystretch{1.3}
\begin{tabular}{l|c|c|c|c|c||c}
\begin{tabular}[c]{@{}c@{}}Cooling\\ type\end{tabular} &
  \begin{tabular}[c]{@{}c@{}}$T_\mathrm{fl}$\\{[$^\circ$C]}\end{tabular} &
  \begin{tabular}[c]{@{}c@{}}$\dot{V}_\mathrm{fl}$\\ {[m$^3$/s]}\end{tabular} &
  \begin{tabular}[c]{@{}c@{}}$P_\mathrm{BTMS}$\\ {[}kW{]}\end{tabular} &
  \begin{tabular}[c]{@{}c@{}}$\Delta E_\mathrm{b}$\\{[}kWh{]}\end{tabular}&
  \begin{tabular}[c]{@{}c@{}}Aircraft\\ weight \\ {[}kg{]}\end{tabular} &
   \begin{tabular}[c]{@{}c@{}}Theoretical \\ flight\\ range {[}km{]}\end{tabular} \\ \hline
None  &   -  & - &  -  & 189.19   & 2345 & \textit{480} \\ \hline
Water & \multirow{2}{*}{24.78} & 3.96$\cdot$10$^{-4}$ & \multirow{2}{*}{9.90} &\multirow{2}{*}{241.12} &\multirow{2}{*}{2732}  &\textit{410}  \\ \cline{1-1} \cline{3-3} \cline{7-7} 
Air   &                   & 6.5$\cdot$10$^{-3}$ &                   &  &  & \textit{373} 
\end{tabular}%
} 
\vspace{-15pt}
\end{table}

\vspace{-5pt}
\subsection{Numerical results}
\begin{figure}[t]
    \centering
    \includegraphics[width=0.85\linewidth]{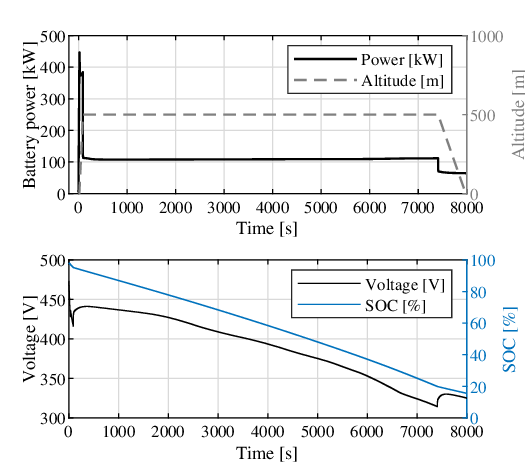}
    \caption{Power demand of aircraft with water-cooled BTMS and flight altitude over time (top), voltage and SOC over flight regime (bottom).}
    \label{fig:Powerdemand}
    \vspace{-15pt}
\end{figure}

\begin{figure}[t]
    \centering
    \includegraphics[width=0.85\linewidth]{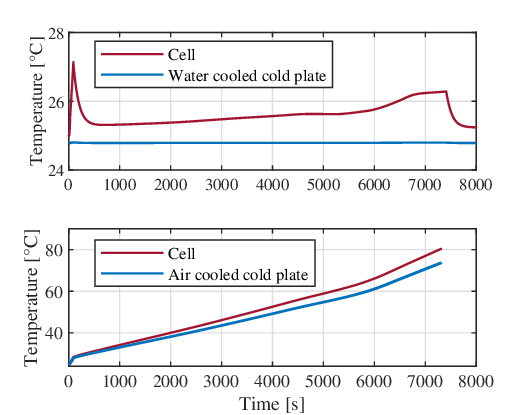}
    \caption{Liquid cooling BTMS (top) and air cooling BTMS (bottom) over flight regime.}
    \label{fig:liq-air}
    \vspace{-10pt}
\end{figure}

We adopt a discrete-time approach, whereby we discretize the model using the trapezoidal method  with a fixed step size of $\Delta t=\unit[1]{s}$. We formulate the problem utilizing MATLAB and conduct the optimization by employing the nonlinear optimization solver \texttt{fmincon} for constrained problems. The aircraft is simulated over a flight cycle with a duration of \unit[8000]{s}, a steady-state flight altitude of \unit[500]{m}, shown in Fig. \ref{fig:Powerdemand}.
Furthermore, we consider the aircraft, battery system, and BTMS specifications of which the optimal design is obtained in the case of water and air cooling. 
 The BTMS assumes a VCM with $K_\mathrm{BTMS}$ = 3 and $\rho_\mathrm{P}$ = 100 W/kg at an ambient temperature of 25$^\circ$C. Table \ref{tab:Results} presents minimum energy consumption values for scenarios without BTMS, with water-cooled BTMS, and with air-cooled BTMS, along with corresponding optimal design variables ($p$). While air cooling achieves similar energy usage to water cooling, it fails to complete the cycle as a result of ineffective heat dissipation, as shown in Fig. \ref{fig:liq-air}. Water cooling maintains cells within their optimal temperature range, as depicted in Fig. \ref{fig:Powerdemand}, adding 16.5\% weight (387 kg) compared to no BTMS.
\begin{figure}[t]
    \centering
    \includegraphics[width=0.85\linewidth]{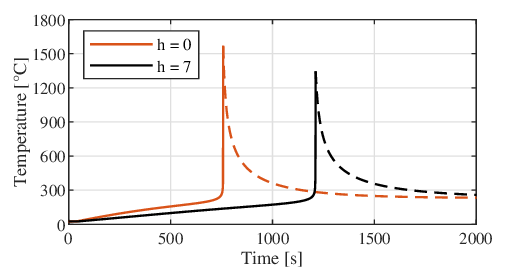}
    \caption{TR predictions of cells for convective heat coefficients h = 0 W/m$^2$/K and h = 7 W/m$^2$/K.}
    \label{fig:TRprediction}
    \vspace{-10pt}
\end{figure}
Next, heating-induced thermal runaway is modeled separately at the cell level.
An 11.84 Ah NMC cell is heated at a rate of 3$^\circ$C/min, corresponding to full power demand. Simulation is conducted at two convection levels: none (h = 0 W/m$^2$/K) and average (h = 7 W/m$^2$/K). Fig. \ref{fig:TRprediction} illustrates cell temperature evolution, with TR trigger points at t = 756 s (h = 0) and t = 1208 s (h = 7). The critical ambient TR trigger temperature exceeds 182$^\circ$C. In Fig. \ref{fig:range}, even without BTMS and convection (h = 0), cell temperatures remain below 86$^\circ$C throughout the flight, indicating no thermal runaway initiation.

%% file: Sections/Conclusion.tex
\section{Conclusion}\label{sec:conclusion}
In this study, we proposed a system-level powertrain and battery sizing approach for light electric passenger aircraft, focusing on battery system design and safety. We developed a parametric framework simulating electrical power consumption across a predefined flight regime, considering thermal and electrical battery dynamics. Investigating the theoretical flight range without BTMS, we noted the impact of varying parallel strings on battery capacity and weight, finding range advantages with increased battery capacity up to the maximum aircraft weight. We proposed a VCM-based BTMS with bottom-mounted cold plates for water and air cooling to control the battery temperature, and its design to achieve minimum energy consumption. Our results showed inadequate heat dissipation and failure to complete the flight regime with air cooling due to constant full power draw, while water cooling meets operational constraints, adding 16.5\% weight to the aircraft. The analysis of heating-induced thermal runaway demonstrated the effectiveness of the battery design in preventing thermal runaway, even without BTMS and convection. Future research may focus on advancements in high-energy-density battery chemistries and integrating HVAC with BTMS to reduce power consumption.